\documentstyle[aps,multicol,epsf]{revtex}

\catcode`\@=11

\def\eqbegin         {  \begin{eqnarray}  }
\def\eqend           {  \end{eqnarray}  }
\def\beq{\begin{equation}}
\def\eeq{\end{equation}}

\def\del          { \partial }

\def\Z{{\bf Z}}

\def\barr{\begin{eqnarray}}
\def\earr{\end{eqnarray}}
\def\u1{\widehat{U(1)}}

\def\lsim{\mathrel{\mathstrut\smash{\ooalign{\raise2.5pt\hbox{$<$}\cr\lower2.5pt\hbox{$\sim$}}}}}
\def\gsim{\mathrel{\mathstrut\smash{\ooalign{\raise2.5pt\hbox{$>$}\cr\lower2.5pt\hbox{$\sim$}}}}}

\title{The Haldane-Rezayi Quantum Hall State and Magnetic Flux}
\author{Kazusumi Ino}
\begin{document}
\begin{center}
\maketitle
{\it Institute for Solid State Physics, University of Tokyo,} \\
{\it Roppongi 7-22-1,  Minatoku,  Tokyo,  106, Japan} 
\end{center}
\begin{abstract}
We consider the general abelian background configurations 
 for the Haldane-Rezayi quantum Hall state. 
We determine the stable configurations to be the ones   
with the spontaneous  flux of $(\Z+1/2) \phi_0$ with $\phi_0 = hc/e$.  
This gives the physical mechanism by which 
 the edge theory of the state becomes identical 
to the one for the 331 state.  
It also provides  a new experimental consequence  which  can be 
tested  in  the enigmatic $\nu=\frac{5}{2}$ plateau in a single layer system. 
\\
PACS: 73.40Hm, 74.20-z,11.25.Hf 
\end{abstract}

\pagenumbering{arabic}
\begin{multicols}{2}
In '87, Willet et al discovered a quantum Hall effect  at the 
even-denominator filling fraction $\nu=5/2$ \cite{willet}. This is so far 
the only even-denominator FQHE found in a single-layer system. While 
tilted field experiments suggest that the ground state of the plateau
 is a spin-unpolarized singlet state \cite{eisen}, 
numerical studies \cite{morf} show that the ground state is a spin-polarized 
pfaffian-like  state \cite{moore}.
This discrepancy has not yet been resolved. 
The $\nu=5/2$ plateau remains an enigma \cite{review,eisen2}.   
 
Soon after the discovery of the $\nu=5/2$ plateau, 
Haldane and Rezayi proposed a variational ansatz for the ground state,
the  so-called  Haldane-Rezayi ( HR ) state.  
\cite{halrez}. It is a spin-singlet paired state.   
The HR state has peculiar physical properties 
such as  the $5p$ degeneracy on the torus \cite{5p,readrezayi} for 
the state with the filling fraction $\nu=1/p$ ($p$ is an even integer). 
In the conformal field theory description, the spin and pairing degrees of 
freedom give  a  non-unitary $c=-2$ conformal field theory 
\cite{milo,wenwu} while the charge degrees of freedom are described by
the $c=1$ chiral boson.  
This theory is known to have physical inconsistency : 
 the failure of the modular invariance for the state on the cylinder. 
There have been some arguments to secure the physical consistency 
\cite{gurarie,inomod,cappelli,gurus}, 
 all  of which propose the edge theory for the HR 
state  to be the same as the one for the Halperin state (331 state) 
\cite{halp}.  
However  no physical mechanim  has been given for this to be hold.

Physically the HR state is a quantum Hall analog of the 
BCS superconductor. 
The pairing symmetry of the paired electrons is $d$-wave. Recently  a
phase transition driven by high magnetic field was discovered  
in  a certain $d$-wave superconductor \cite{krishana}. Laughlin  proposed that 
 the phase transition  is a consequence of the development of 
 $T$-violating  order parameter \cite{laughd-wave}. 
The essential point of his arguments is the spontaneous 
induction of magnetic moment.  
There are some other discussions on this pairing order \cite{bal}.

As the HR state is a quantum Hall and thus $T$-violating
 analog of $d$-wave superconductor, Laughlin's 
argument suggests the possibility of induced  moment or flux 
for the state.  In this paper, we study the possible background 
configurations of the HR state.  We  show that there is indeed spontaneous 
induction of flux in the  most stable configurations. 
The modular invariance is restored in these configurations.
We also study   the bulk excitations by a variant of 
Laughlin's thought experiment and 
determine the composite laws of the excitations. 

Let us first consider the edge conformal field theory of the HR state on the 
disk.  The conformal field theory can be deduced by studying the wave function 
\cite{wenwu,milo}. The conformal field theory for the spin and pairing degrees 
of freedom is a   $c=-2$ conformal field theory, while the 
charge degrees of freedom are described by the 
 $c=1$ chiral boson $\varphi$. In the free field representation, 
the $c=-2$  theory is realized by two anticommuting scalars
(symplectic fermions )$\Psi^{\uparrow}$,
$\Psi^{\downarrow}$ \cite{kausch,milo}. 
In the $ \nu=1/p$ state, 
the electron fields are given by 
$\del\Psi^{\uparrow(\downarrow)}e^{i\sqrt{p}\varphi}$.
They enlarge the chiral algebra of the state. 
The zero modes do not enter to the chiral algebra.
 The edge excitations are determined by the no 
monodromy condition with the electron fields.  We treat the zero modes separately,  then  
 the primary fields for  each sector  are given by 
\eqbegin
Ie^{i\frac{r}{\sqrt{p}}\varphi}, \hspace{1.5mm} 
\del\Psi^{\uparrow(\downarrow)}e^{i\frac{r}{\sqrt{p}}\varphi},
\hspace{1.5mm}
\sigma e^{i\frac{r}{2\sqrt{p}}\varphi},
\hspace{1.5mm}
\widetilde{\sigma}^{\alpha}e^{i\frac{r}{2\sqrt{p}}\varphi},
\label{excitations}
\eqend
where $I$ is the identity and 
$\sigma,\sigma^{\alpha}$ are the spin and disorder field respectively, 
creating twisted sectors for $\del\Psi^{\uparrow(\downarrow)}$. 
The fields  also correspond to the quasiparticles 
in the bulk. 
The edge excitations in each sector are generated by acting 
the Virasoro algebra on the corresponding  primary field.
The complete description of edge excitations is conveniently summarized in the
 Virasoro characters. For the fermionic excitations, they are
\eqbegin 
\chi_{I}(\tau)
&=&\frac{1}{2}
\left(\frac{\vartheta_2(\tau)}{2\eta(\tau)}+\eta(\tau)^2\right),
\\
\chi_{\Psi}(\tau) &=& \frac{1}{2}
\left(\frac{\vartheta_2(\tau)}{2\eta(\tau)}-\eta(\tau)^2\right),  
\\
\chi_{\sigma}(\tau) 
&=& \frac{1}{2}\left(\frac{\vartheta_3(\tau)}{\eta(\tau)}
+\frac{\vartheta_4(\tau)}{\eta(\tau)}\right), 
\\ 
\chi_{\widetilde{\sigma}}(\tau)
&=& \frac{1}{2}\left(\frac{\vartheta_3(\tau)}{\eta(\tau)}
 -\frac{\vartheta_4(\tau)}{\eta(\tau)}\right),  
\label{c=-2chara}
\eqend
while the characters for the U(1) sector are given by  
$\chi_{p}(\tau) = \frac{\vartheta_3(\nu\tau)}{\eta(\tau)},\hspace{3mm}
\chi^{(1/2)}_{p}(\tau) =  \frac{\vartheta_2(\nu\tau)}{\eta(\tau)}$.
Here  $\vartheta$'s are the Jacobi theta functions and 
$\eta$ is the Dedekind function 
$\eta(\tau)=q^{\frac{1}{24}}\prod_{n=1}^{\infty}(1-q^n)$ with 
$q=e^{2\pi i \tau}$ 
. 
By taking $\tau =iT_0/k_B T$ where $k_B$ is the Boltzmann constant 
(we set $k_B=1$ below) and 
$T_0$ is the level spacing of the system, 
combinations of these Virasoro characters  
represent contributions from each sector to the grand 
partition function.  By summing up these terms, 
 the grand partition function becomes
\eqbegin 
{\cal Z}_{\rm HR}(T) = \frac{1}{\eta^2}
\bigl[\vartheta_2(\tau)\vartheta_3(\nu\tau)
+\vartheta_3(\tau)\vartheta_2(\nu\tau)\bigr]
\eqend 
where we have included the fermionic zero modes.

We now consider the general abelian background configurations of the state. 
The system has a U(1) symmetry generated by $J^c=
\frac{i}{\sqrt{p}}\del\varphi$. 
We will denote this U(1) symmetry as ${\rm U(1)}_c$.  
The background configuration coupling to the U(1) current 
is the magnetic flux $\Phi_c$.  
There is  also another U(1) symmetry in this system which we will denote 
as ${\rm U(1)}_{s}$. It is twice the projected $(z$-component) spin current 
$J^s= \Psi^{\uparrow}\del\Psi^{\uparrow}-\Psi^{\downarrow}\del\Psi^{\downarrow}$.
One can also couple magnetic flux $\Phi_s$ to this current.  
With the introduction of $\Phi_c$ and $\Phi_s$, the HR state is parametrized 
by the two dimensional space ($\Phi_c,\Phi_s$). Under the presense of 
the fluxes $\Phi_c$ and $\Phi_s$,  there are spectral flows 
of the edge excitations.  For example, $\chi_p$ changes as 
$\chi_p(\tau) \rightarrow \chi_p(\tau,\Phi_c)= 
\sqrt{p}\frac{\vartheta_3(\Phi_c|t)}{\eta(t)}$
where $t=-1/\tau = ik_BT/T_0$ and we measure $\Phi_c$  with 
the unit $\phi_0 = hc/e$. 
After some modular transformations, we end up with the grand partition
function for the 
$(\Phi_c,\Phi_s)$ HR state as 
\eqbegin 
{\cal Z}_{\rm HR} = \frac{\sqrt{p}}{\eta(t)^2}\bigl[(\vartheta_4(\Phi_s|t)
\vartheta_3(\Phi_c|pt) + \vartheta_3(\Phi_s|t)\vartheta_4(\Phi_s|pt) \bigr]. 
\eqend
${\cal Z}_{\rm HR}(T,\Phi_c,\Phi_s)$ has periods $1$ for both $\Phi_c$
and $\Phi_s$.
The free energy for the edge excitations of the HR state becomes 
$ {\cal F}_{\rm HR}(T, \Phi_c,\Phi_s)  
= - k_B T {\rm ln} {\cal Z}_{\rm HR}$.

Now we'd like to determine the stable configurations. 
We investigate the behavior of ${\cal F}_{\rm HR}(T,\Phi_c,\Phi_s)$ on the 
$(\Phi_c,\Phi_s)$ plane.  
Let us first consider the behavior at $T \gsim T_0$.  
The minima are determined by the minimum conditions 
$j = \frac{\del{\cal F}_{\rm HR}}{\del \Phi} = 0, \hspace{5mm} 
\frac{\del^2{\cal F}_{\rm HR}}{\del \Phi^{\alpha}\del \Phi^{\beta}}
\delta\Phi^\alpha\delta\Phi^\beta > 0.
$
Here $j$ is the induced persistent edge current for $\Phi_c$ \cite{ino}. 
By solving these conditions, the space of the minima is
 determined to be  
\eqbegin 
\Gamma = \{(x,y)| (x,y) \in (\Z/2,\Z/2),   
x+y \in \Z+1/2 \}. 
\eqend 
The lattice structure of $\Gamma$ is due to an extra periodicity of 
${\cal Z}_{\rm HR}$  with period $(\frac{1}{2},\frac{1}{2})$.
Thus the stable background configurations for the HR
state are   not the ones with no  flux or integer flux, 
but the ones with  the  half integer net  flux. 
Actually  the integer  net flux lattice  is the set of {\it maxima} for
$T \gsim T_0$.  
We also note that the unitarity of the edge theory
 is  restored in these configurations.   
Thus we conclude that the HR state is unstable by itself, and 
it has to generate the half integer flux for either $\Phi_c$ or $\Phi_s$ 
 {\it spontaneously}. The half integer flux is physically observable, 
while the integer flux is not physically observable.

\begin{figure}
\hspace{3.3in}
\epsfbox{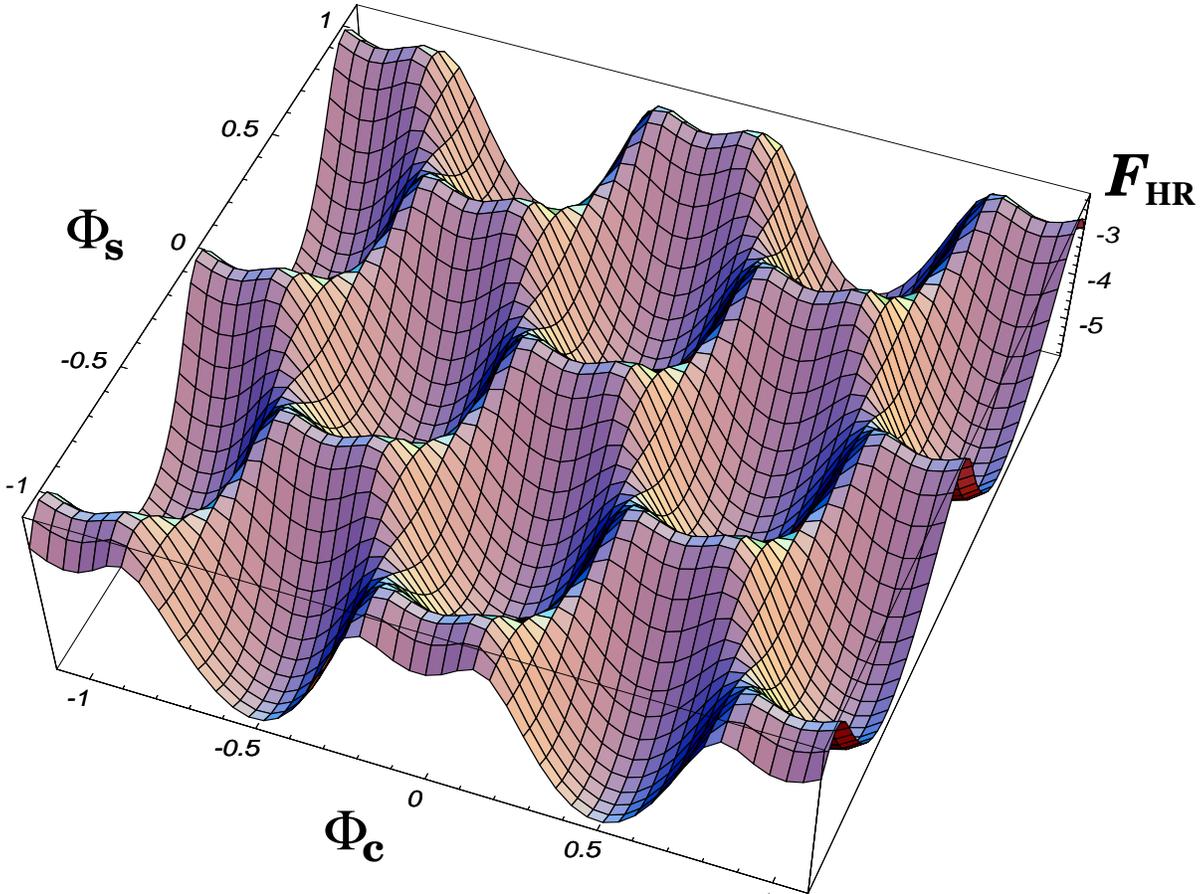}
\caption[Lattice]
{Free energy for the cylinder 
as function of $\Phi_c$ and $\Phi_s$ at $T/T_0 = 1.36$ ($p=2$). 
The terms not dependent on flux are omitted. }
\label{free}
\end{figure}

Let us next consider the HR state on the cylinder. 
We restrict the excitations to the edge excitations. Then there are 
constraints between the excitations on two edges. 
The {\it non-modular invariant} cylinder partition function 
was obtained in Ref.\cite {milo}. 
It is the sum of the combinations of  the characters for the left
 and right movers under the constraints between two edges. 
The partition function with the flux $\Phi_c$ and its 
modular behavior were studied in \cite{inomod}.
Its extension to include $\Phi_s$ is readily done.   
It can be written in terms of theta functions as 
\end{multicols}
\eqbegin
{\cal Z}^{\rm cyl}_{\rm HR}= \frac{1}{2p |\eta(t)|^4} 
\sum_{r=1}^{p} \{ \biggl|\vartheta_4(\Phi_s|t)\vartheta_3(\frac{\Phi_c+r}{p}|t/p)\biggr|^2
+ \biggl|\vartheta_3(\Phi_s|t)\vartheta_3(\frac{\Phi_c+r+1/2}{p}|t/p)\biggr|^2
\nonumber 
\\ 
+\biggl|\vartheta_1(\Phi_s|t)\vartheta_2(\frac{\Phi_c+r}{p}|t/p)\biggr|^2 
+\biggl|\vartheta_2(\Phi_s|t)\vartheta_2(\frac{\Phi_c+r+1/2}{p}|t/p)\biggr|^2 \}.
\label{HRcyl}
\eqend
\begin{multicols}{2}
${\cal Z}^{\rm cyl}_{\rm HR}$ has the same periodicity in
$(\Phi_c,\Phi_s)$  as ${\cal Z}_{\rm HR}$. 
We plot the free energy as a function of $(\Phi_c,\Phi_s)$ in
Fig.\ref{free}. 
By using the explicit form above,  it is shown that 
the free energy 
${\cal F}^{\rm cyl}_{\rm HR} = -k_B T {\rm ln} {\cal Z}^{\rm cyl}_{\rm HR}$ shows 
a similar behavior with ${\cal F}_{\rm HR}$ and  shares 
the same space of minima $\Gamma$ and the space of maxima. 
Thus there is spontaneously generated induced 
flux along the center of the cylinder.

Let us next consider the modular behavior of the partition function 
${\cal Z}^{\rm cyl}_{\rm HR}$.
By using the analogous methods used in Ref.\cite{inomod}, it is shown that 
there is a relation between the Haldane-Rezayi state and 
the Halperin 331 state 
\eqbegin
{\cal Z}^{\rm cyl}_{\rm HR}(T,x,y) = {\cal Z}^{\rm cyl}_{331}(T), 
\hspace{2mm}(x,y)\in \Gamma, 
\eqend
where $ {\cal Z}^{\rm cyl}_{331}(T)$  is the cylinder partition
function for the 331 state. 
The identity generalizes the previous results of Refs.\cite{inomod,cappelli}.
As ${\cal Z}_{331}(T)$ is modular invariant, modular invariance is restored 
on the minimum free-energy configurations. 
  The failure of the modular
invariance of (\ref{HRcyl}) occurs in its non-minimum free-energy configurations. 
Thus the generation of flux gives the physical mechanism that makes
the edge theory of the Haldane-Rezayi state to be identical to the one for 
the 331 state.

We next consider the temperature region enough below the level spacing $T_0$.
 In this case it is shown that ${\cal Z}_{\rm HR}$  has  an 
 enlarged space of minima $ 
\Gamma_0 = \{(x,y)| (x,y) \in (\Z/2 ,\Z/2)  \}$.
 This property results in the $1/2$ period in the induced edge current 
 at zero temperature \cite{ino}. One can also prove   the same behavior for 
${\cal Z}^{\rm  cyl}_{\rm HR}$. $\Gamma$ remains the set of the most
stable configurations also in this case.

We now turn to study the bulk excitations of the HR state 
by refining Laughlin's thought experiment  
on quantum Hall state on the cylinder in terms of the edge partition 
function \cite{laugh}.

Let us consider the Laughlin state with $\nu=1/p$ on the cylinder 
without flux.  The edge state are in a minimum of 
the free energy ${\cal F}_L=-k_BT{\rm ln}Z_{L}$ where  
$Z_L = \frac{1}{p\eta(t)^2}\sum_{1}^{p}[\vartheta_3(\frac{\Phi_c+r}{p}| t/p)^2]$. 
Imagine that we turn on the unit flux  through 
the cylinder.  This causes the transportation of the elementary quasihole 
$e^{i\frac{1}{\sqrt{p}}\varphi}$ from one edge to the other. 
Indeed the flux $\Phi_c = 1 $ is another 
minimum of ${\cal F}_L$.  Thus the elementary quasihole is 
the excitation which corresponds to the transition 
between the nearest minima of the free energy of the edge excitation.
The successive transitions between minima 
caused by  turning on  another flux
give the $\Z$  structure of the bulk excitations, 
which is nothing but the fusion rules of the bulk CFT.

\begin{figure}
\hspace{2.0 in}
\epsfbox{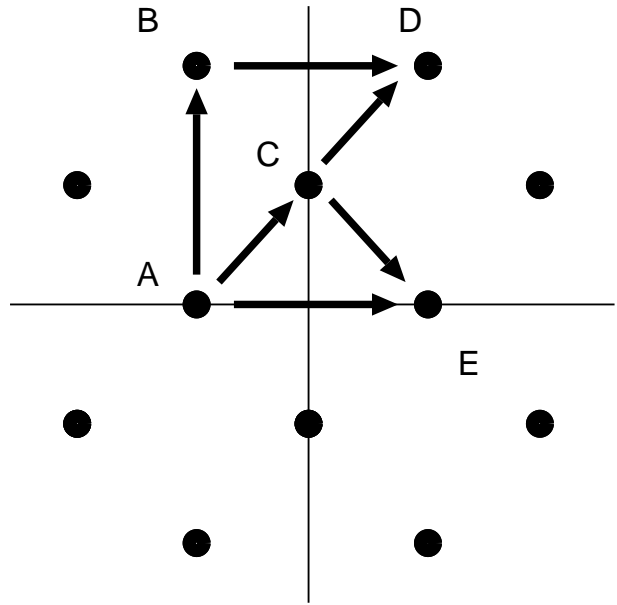}
\caption[transition]
{The lattice of minima of ${\cal F}_{\rm HR}$  at $T \gsim T_0$ and 
transitions among the stable configurations.}
\label{trans}
\end{figure}

We generalize this argument to the HR state to determine the
bulk excitation structure. When the temperature is far below 
the level spacing, the transition between minima becomes
difficult.  So we  consider the region $T \gsim T_0$.  
First consider the HR state with flux A $(-\frac{1}{2},0)$ 
(Fig.\ref{trans}). 
Imagine that we turn on the flux $\delta\Phi_c = 1$. 
This causes a transition from A to another minimum E  
$(\frac{1}{2},0)$. This is the same transition as 
in the Laughlin case. The Laughlin quasihole $e^{i\frac{1}{\sqrt{p}}\varphi}$ 
is transported from one edge to the other.   

Next imagine that we turn on the flux $\delta\Phi_c = \frac{1}{2}, 
\delta\Phi_s = \frac{1}{2}$. This causes a transition from 
A to C $(0,\frac{1}{2})$. From the changes in the U(1) 
quantum numbers on two edges, the bulk excitation transported is 
$\sigma^{\frac{\alpha}{2}}e^{i\frac{1}{2\sqrt{p}}\varphi}$ 
where $\sigma^{\frac{\alpha}{2}}$  is   the ${\rm U(1)}_s$ charge
$\frac{1}{2}$  component of spin field
$\sigma$.   
As we see in Fig.\ref{free}, this 
transition corresponds to  the valley between the minima and 
energetically most favorable. Therefore the corresponding bulk excitation is 
the lightest quasiparticle in this system. 

Next imagine that we turn on the flux $\delta\Phi_s =1$.
This causes a transition from A to B $(-\frac{1}{2},1)$. 
The corresponding bulk excitation should have the ${\rm U(1)}_s$ 
charge $1$. For example $\del\Psi^{\alpha}$ has 
${\rm U(1)}$ charge $1$, but this field is not suitable to be 
transported  because this transition should map the sectors of 
the Hilbert space ${\cal H}_{m,n}$ to ${\cal H}_{m,n+1}$ by one-to-one 
($(m,n)$: ${\rm U(1)}_c \times {\rm U(1)}_s$ charges). 
As the fermionic edge excitations have the zero modes, 
the map must involve the fermionic zero modes.  Thus the appropriate 
field for  the excitation which accompanies the transition A to B is 
$\Psi^{\alpha}$.

Let us next consider the successive transitions  $
A \rightarrow C \rightarrow D, \hspace{2mm}
A \rightarrow C \rightarrow E$ (Fig.\ref{trans}). 
In these transitions, two cases are possible. One is the 
successive occurrences of the transitions in order. In this case, 
two bulk excitations are  created in order. 
The other is the simultaneous occurrence of the two transitions.     
The two bulk excitations will form  an excitation corresponding to 
$ A \rightarrow D $ or $A \rightarrow E$. These cases  give 
the same consequence, thus physically equivalent.

One can relate the first case to the product or composition $\times$ 
of the two quasiparticles, 
which is {\it fusing} in the language of conformal field theory. 
Also the latter case gives the resulting representations of 
chiral algebra by the composition of the two quasiparticles.

The transition $A \rightarrow C$ and $C \rightarrow D$ are 
 caused by the transportation of 
the quasihole $\sigma^{\frac{\alpha}{2}}e^{i\frac{1}{2\sqrt{p}}\varphi}$ 
from one edge to the other, while $C \rightarrow E$ is from  the  
transportation of the quasihole 
$\sigma^{-\frac{\alpha}{2}}e^{i\frac{1}{2\sqrt{p}}\varphi}$.  
On the other hand, $A \rightarrow D$ and $A \rightarrow E$ 
make the transportation 
of $I e^{i\frac{1}{\sqrt{p}}\varphi}$ and 
$\Psi^{\alpha}e^{i\frac{1}{\sqrt{p}}\varphi}$ respectively.
As these process are physically equivalent, we have thus 
the following identities: 
\eqbegin  
\sigma^{\frac{\alpha}{2}}e^{i\frac{1}{2\sqrt{p}}\varphi}\times 
\sigma^{\frac{\alpha}{2}}e^{i\frac{1}{2\sqrt{p}}\varphi} \sim 
\Psi^{\alpha}e^{i\frac{1}{\sqrt{p}}\varphi}, \\
\sigma^{\frac{\alpha}{2}}e^{i\frac{1}{2\sqrt{p}}\varphi}\times 
\sigma^{-\frac{\alpha}{2}}e^{i\frac{1}{2\sqrt{p}}\varphi} \sim 
I e^{i\frac{1}{\sqrt{p}}\varphi}.
\eqend 
By combining these identities and omitting the charge 
part , we have the following identity 
for the neutral spin field $\sigma$, $ 
\sigma \times \sigma \sim  I + \widetilde{I}$ 
where $\widetilde{I} = \epsilon_{\alpha\beta}\Psi^{\alpha}
\Psi^{\beta}$ with 
$\epsilon_{\uparrow\downarrow}=-\epsilon_{\downarrow\uparrow}= 1/2$
 and zero otherwise.
This is  nothing but the fusion rules of the spin field and 
the logarithmic operator in the $c = -2$ {\it logarithmic} CFT \cite{logCFT}. 
Other fusion rules are deduced 
similarly.  
As the chiral algebra of the bulk does not include 
the fermionic zero modes and only has  $\del\Psi
e^{i\frac{1}{\sqrt{p}}\varphi}$ 
as extending field, the field $\widetilde{I}e^{i\frac{1}{\sqrt{p}}\varphi}$ 
cannot be generated from the fields in (\ref{excitations}).  
Thus $\widetilde{I}e^{i\frac{1}{\sqrt{p}}\varphi}$ 
results in a new sector. It leads to the 
$5p$ sectors in the bulk, thus  
 the  $5p$ degeneracy of the HR state on the torus. 
On the other hand, in the edge theory where 
$\Psi$ effectively has a half integer moding by the 
presence of half integer flux, $\widetilde{I}$ does not lead to a new sector, 
which results in  the $4p$ sectors. 

Thus we have determined the structure of the bulk excitations of the HR state 
from the space of minima of the edge free energy. There have been some
arguments on other possibilities on the bulk theory of the HR state  
\cite{lee,cappelli}. 
Our discussion shows  that  
the bulk  of the HR state 
is consistently described by the $c = -2$ logarithmic CFT \cite{gurarie}.


Let us next discuss implications of our result for the $\nu=5/2$ plateau. 
Recent numerical studies \cite{morf} for the small electron system on
the sphere show that 
the $\nu= \frac{5}{2} $ state may be  spin-polarized, 
while a spin-unpolarized state is not favored by its 
relatively higher energy per one electron, and the energy of 
the system is scales as $O(N)$ where $N$ is the number of electrons.
Our result shows that the generation of flux 
in  the HR state in a finite system with edges 
 lowers the energy of the its edge state.  However  its reduction 
 only scales as $O(1)$. 
 Thus, in the thermodynamic limit, 
 our result does not by itself implies the bulk state of 
the $\nu=\frac{5}{2}$ plateau to be the HR state with the stable background.

On the other hand, experimental studies \cite{willet,eisen}  
indeed  favor a spin-unpolarized singlet state. 
The results of the tilted field experiment 
are naturally explained by the HR state. 
Our result provides a new experimetal test on this. 
If the state is the HR state,  then 
for the sample with  shape like the annulus, the flux through the center 
 should be  quantized as  $(\Z+\frac{1}{2})\phi_0$. 
This is a unique property for the HR state and its detection 
in the $\nu=5/2$ plateau  gives  an clear evidence that the plateau
 realizes the Haldane-Rezayi state.

Finally the physical mechanism we describe in this paper   
suggests  the possibility  of larger class of vacua in the FQH systems, 
which we will discuss in detail elsewhere \cite{backtofuture}.

{\it Acknowledgement  } 
We would like to thank Mahito Kohmoto and  
Jun'ichi Shiraishi for discussions.

\def\NP{{ Nucl. Phys.\ }}
\def\PRL{{ Phys. Rev. Lett.\ }}
\def\PL{{ Phys. Lett.\ }}
\def\PR{{ Phys. Rev.\ }}
\def\CMP{{ Comm. Math. Phys.\ }}
\def\IJMP{{ Int. J. Mod. Phys.\ }}
\def\MPL{{ Mod. Phys. Lett.\ }}
\def\RMP{{ Rev. Mod. Phys.\ }}
\def\AP{{  Ann. Phys. (NY)\ }}

\end{multicols}

\end{document}